\documentclass[pre,aps,onecolumn,superscriptaddress]{revtex4-1}
\usepackage{amsmath,amssymb,graphicx}
\usepackage{epsfig}
\usepackage{graphicx}
\usepackage{textcomp}

\begin{document}

\title{Extreme Current Fluctuations in a Nonstationary Stochastic Heat Flow}

\author{Baruch Meerson}
\email{meerson@mail.huji.ac.il}
\affiliation{Racah Institute of Physics, Hebrew University of
Jerusalem, Jerusalem 91904, Israel}
\author{Pavel V. Sasorov}
\email{pavel.sasorov@gmail.com}
\affiliation{Keldysh Institute of Applied Mathematics, Moscow 125047, Russia}

\begin{abstract}
\noindent
We employ the Hamiltonian formalism of macroscopic fluctuation theory to study large deviations of integrated current  in the Kipnis--Marchioro–-Presutti (KMP) model of stochastic heat flow when starting from a step-like initial condition. The KMP model belongs to the hyperbolic universality class where diffusion remains relevant no matter how large the fluctuating current is. The extreme current statistics for the KMP model turns out to be sub-Gaussian, as distinguished from the super-Gaussian statistics found for the Symmetric Simple Exclusion Process and other models of the elliptic class.  The most probable time history of the system, which dominates the extreme current statistics of the KMP model, involves two large-amplitude solitary pulses: of the energy density field and of the conjugate ``momentum" field. The coupled pulses propagate with a constant speed, but their amplitudes slowly grow with time, as the energy density pulse collects most of the available energy on its way.

\end{abstract}
\maketitle

\tableofcontents
\nopagebreak

\section{Introduction}
\label{intro}
Large fluctuations of currents of matter or energy in many-particle systems far from 
thermodynamic equilibrium have attracted much recent interest. Important progress has
been achieved in studies of large fluctuations in boundary-driven nonequilibrium steady
states (NESS) of stochastic diffusive lattice gases, see e.g. \cite{D07} and references therein.
Diffusive
lattice gases constitute a broad family of classical
transport models which provide a succinct representation of
different aspects of matter and energy transport  in extended systems \cite{KL99,Spohn,L99}.
Among the most extensively studied lattice gas models are the Symmetric Simple
Exclusion Process (SSEP) \cite{Spohn,KL99,L99,SZ95,D98,S00,BE07,KRB10} and the Kipnis--Marchioro–-Presutti (KMP) model \cite{KMP,BGL}.
In the SSEP there is at most one particle on a lattice site, and a particle can randomly hop to a neighboring site if that site is empty. If it is occupied by another particle, nothing happens. The SSEP accounts, in a simple way, for  inter-particle repulsion. The KMP model involves a lattice of mechanically uncoupled harmonic oscillators which randomly redistribute energy among neighbors. The KMP model was devised to mimic heat conduction in a crystal. Here, because of the \textit{a priori} stochastic dynamics, the validity of Fourier's law of heat conduction is proven rigorously \cite{KMP}.

The large deviation function \cite{LDF} of the current in the NESS has been intensively studied in the  last decade, see reviews \cite{D07,Jona}. It has been found that, for some diffusive lattice gases, such as the SSEP, the large deviation function of the current is Gaussian and universal \cite{D07,Jona}. For other gases, such as that described by the KMP model, the Gaussian statistics give way, at sufficiently large currents, to different statistics. This change occurs via a second order phase transition as the system length tends to infinity. This phase transition is intimately related to the fact that the optimal density profile, conditional on the given current, changes from time independent to time dependent  \cite{Bertini05,Bodineau,Hurtado}. Here we will deal with \emph{non-stationary} fluctuations of diffusive lattice gases which are much
less understood \cite{DG2009a,DG2009b,van,KM_var,void,varadhan,MS2013}. As Refs. \cite{DG2009a,DG2009b,KM_var,varadhan,MS2013}, we will consider a diffusive
lattice gas on an infinite line and study fluctuations
of integrated current $J$ through the origin $x=0$ during a specified time $T$,
when starting from a deterministic step-like density (or energy density) profile
\begin{equation}
\label{step}
n(x,t=0) =
\begin{cases}
n_-,   & x<0,\\
n_+,  & x>0.
\end{cases}
\end{equation}
If the shot noise is neglected, the large-scale dynamics of diffusive lattice gases is described by
the diffusion equation
\begin{equation}\label{diffusion}
\partial_t n = \partial_x \!\left[D(n) \,\partial_x n\right],
\end{equation}
where $D(n)$ is the diffusion coefficient. Having solved this equation with the initial condition \eqref{step}, one can determine the \emph{average} integrated current
at time $T$:
\begin{equation}\label{averageJ}
\langle J(T)\rangle = \int_0^\infty dx\, [n(x,T)-n_+].
\end{equation}
The actual current $J$ fluctuates due to the shot noise of the microscopic model. It is interesting to determine the probability density ${\cal P}(J,T)$ of these fluctuations which, at large $T$, exhibits scaling behavior. The analysis of the large-$T$ limit is facilitated by the fact that a broad class of lattice gases can be described in this limit by the \emph{fluctuating hydrodynamics} which involves a
Langevin equation \cite{Spohn,KL99}:
\begin{equation}
\label{Lang}
     \partial_t n = \partial_{x} [D(n)\, \partial_x n] +\partial_x \left[\sqrt{\sigma(n)} \,\eta(x,t)\right].
\end{equation}
Here $\eta(x,t)$ is a Gaussian noise with zero average, which is delta-correlated both in space and in time:
\begin{equation}
\left\langle \eta(x,t)\eta(x_1,t_1))\right\rangle=\delta(x-x_1)\, \delta(t-t_1),
\label{deltacorr}
\end{equation}
and the brackets denote ensemble averaging. At the level of fluctuating hydrodynamics a diffusive lattice gas is fully characterized by the diffusion coefficient $D(n)$ and another coefficient, $\sigma(n)$, which comes from the shot noise and is equal to twice the mobility of the gas \cite{Jona}. These coefficients obey an analog of the Einstein relation: $F^{\prime\prime}(n)=2 D(n)/\sigma(n)$, where $F(n)$ is the equilibrium free energy of the homogeneous system \cite{Spohn,Jona,KM_var}, and primes denote derivatives with respect to the (single) argument. For the SSEP $D=1$ and $\sigma(n)=2n(1-n)$, whereas for the KMP $D=1$ and $\sigma(n)=2n^2$ \cite{4n2}.

The next level of theory, \emph{macroscopic fluctuation theory} (MFT), is well suited for dealing with large deviations, as it exploits the \emph{typical} noise strength in the system  as a small parameter (the latter scales as $1/\sqrt{N}$ where $N$ is the typical number of particles in the relevant region of space).  Originally derived for the NESS \cite{Bertini,Bertini05}, the MFT was then extended to non-stationary settings including the large current statistics with a step-like initial density profile \cite{DG2009b}. The MFT can be formulated as a classical Hamiltonian field theory \cite{Bertini,DG2009b,Tailleur}, and we will adopt the Hamiltonian approach in the following. A celebrated related theory in physics literature is the (weak noise limit of) Martin-Siggia-Rose field-theoretical formalism  for continuous stochastic systems \cite{MSR} which has been employed in numerous works.

The problem of complete statistics of integrated current was formulated within the framework of MFT in Ref. \cite{DG2009b}.
In the same Ref. \cite{DG2009b} the problem was solved for non-interacting random walkers. More recently, the probability of observing small fluctuations of current around the mean has been found for diffusive lattice gases with $D(q)=1$ and an arbitrary $\sigma(q)$ \cite{KM_var}, see also Ref. \cite{varadhan}. We also mention here, for completeness, the quite different \emph{annealed} setting,
when the initial density (or energy density) field at $x<0$ (correspondingly, at $x>0$) is allowed to fluctuate and chosen from the equilibrium probability distribution corresponding to density $n_{-}$ (correspondingly, $n_{+}$). In Ref. \cite{DG2009a} the complete statistics of the integrated current was found, for the annealed setting, for the SSEP. This was achieved by exactly solving the microscopic model. Then it was shown, in the framework of the MFT, that the complete statistics of integrated current for the KMP model (again, in the annealed setting) can be expressed through the corresponding statistics for the SSEP \cite{DG2009b}.

Now let us return to the deterministic (or quenched) initial condition. In the absence of a complete solution for interacting particles,  it is of great interest to at least determine the extreme current asymptotic ${\cal P}(J\to \infty,T)$ and see whether it is universal for different diffusive lattice gases.   For the random walkers, this asymptotic has the form \cite{DG2009b,MS2013}
\begin{equation}\label{PRW}
\ln {\cal P}(J\to \infty,T)\simeq - \frac{J^3}{12 n_{-}^2 T}.
\end{equation}
Derrida and Gerschenfeld \cite{DG2009b} conjectured that the super-Gaussian decay $\sim J^3/T$  holds for a whole class of \emph{interacting} gases, and proved their conjecture for an $n$-independent $D$
and $\sigma(n)\leq n+\text{const}$ for $0\leq n\leq n_{\text{max}}$, and $\sigma(n)=0$ otherwise.

Further progress has been made in our recent work \cite{MS2013}, where two major universality classes of diffusive lattice gases with respect to the extreme current statistics -- the elliptic and hyperbolic classes --  have been identified. They are determined by the sign of the second derivative $\sigma^{\prime\prime}(n)$.  For the elliptic class, $\sigma^{\prime\prime}(n)<0$, the Derrida--Gerschenfeld conjecture holds \cite{MS2013}, as one finds
\begin{equation}\label{Pgeneric}
\ln {\cal P}(J\to \infty,T)\simeq -\frac{f(n_{-},n_+) J^3}{T}.
\end{equation}
Furthermore, for the SSEP [actually, for any model with $D(n)=1$ and $\sigma(n)=an-bn^2$, where $a>0$ and $b>0$] the function $f(n_-,n_+)$ can be found analytically \cite{MS2013,MSV2013}.
This progress was possible because, at large $J$, the diffusion terms in the MFT equations (see below) can be neglected compared with the terms due to fluctuations.  Furthermore, the resulting reduced MFT equations can be mapped into inviscid hydrodynamics of an effective fluid with a negative compressibility, ensuing an elliptic flow \cite{MS2013}.  For $\sigma(n)=an-bn^2$ the ensuing mathematical problem is exactly soluble. The complete solution also include  regions where the effective fluid density vanishes, and shock discontinuities develop, but these do not contribute to $\ln {\cal P}(J\to \infty,T)$ \cite{MS2013,MSV2013}. 

The present work focuses on the KMP model. Here $\sigma^{\prime\prime}(n)=4>0$, so this model belongs to the hyperbolic class. Here the inviscid equations, obtained when neglecting diffusion, turn out to have a singularity at all times. To understand the character of this singularity, and how to properly include it in an inviscid hydrodynamic description, we have to return to the complete MFT equations where the singularity is regularized by diffusion. We show here that, for the KMP model, the regularized singularity has the form of two coupled large-amplitude solitary pulses: of the density field and of the conjugate ``momentum" field which appears in the Hamiltonian formulation of the MFT. In the large-current limit the coupled pulses travel with constant speed, but their amplitudes steadily grow with time, as the energy density pulse collects most of the available energy on its way. The physical mechanism behind the transport of (an unusually large amount of) energy in the form of a single pulse can be tracked down to the fact that, in the KMP model, $\sigma(n)$ grows without limit, and sufficiently fast, with $n$. As a result, the system utilizes its noise in the most efficient way if the transported energy is clumped into a short pulse. Importantly, the dominant contribution to $\ln P(J\to \infty,T)$ comes from the regularized singularity, rather than from the large inviscid flow regions. As we show here, this leads to a sub-Gaussian extreme current statistics:
\begin{equation}
 \ln {\cal P}(J \to \infty,T) = -\sqrt{T} j\left[2 \ln j+\ln (2 \ln j)+\frac{\ln(2 \ln j)}{2 \ln j}+\dots\right], \;\;\;\;\;
 j=\frac{J}{n_- \sqrt{T}}, \label{KMPexpansion}
\end{equation}
as distinguished from Eq.~(\ref{Pgeneric}). A salient feature of Eq.~(\ref{KMPexpansion}) is its independence of $n_{+}$. The reason for it will become clear as we proceed.

Here is a plan of the remainder of the paper. In section \ref{recap} we recap the MFT formulation of the extreme current statistics problem for any $D(q)$ and $\sigma(q)$ \cite{DG2009b,KM_var} and emphasize the role of three conservation laws in the problem. In Section \ref{singul} we show that the inviscid formulation
of the problem for the KMP model necessarily includes a singularity at all times.  Section \ref{exactsol} presents exact traveling pulse solutions of the full MFT equations for the KMP model. These play a pivotal
 role in section
\ref{perturb} where we develop a perturbation theory for the the extreme current statistics problem for the KMP model.  We briefly summarize and discuss our results in Section \ref{discus}.

\section{Macroscopic Fluctuation Theory of Integrated Current}
\label{recap}
A specified integrated current $J$ at time $T$ is described by the equation
\begin{equation}
\label{current0}
\int_0^\infty dx\, [q(x,t=1)-n_{+}] =J/\sqrt{T}= \tilde{j},
\end{equation}
where we have rescaled $t$ and $x$ by $T$ and $\sqrt{T}$, respectively. The density (or energy density) field $q(x,t)$ and the canonically conjugate ``momentum" field $p(x,t)$ obey Hamilton equations
\begin{eqnarray}
\partial_t q &=& \partial_x \left[D(q)\, \partial_x q\right]
-  \partial_x \left[\sigma(q)\, \partial_x p\right],\label{q:eqfull}\\
\partial_t p &=& - D(q) \partial_{x}^2 p
- \frac{1}{2}\sigma^{\prime}(q)\!\left(\partial_x p\right)^2,\label{p:eqfull}
\end{eqnarray}
which emerge as saddle-point equations of the field theory corresponding to the Langevin equation (\ref{Lang}). The Hamiltonian functional is $H=\int_{-\infty}^\infty dx\,h$, where
\begin{equation}\label{hath}
 h = -D(q)\, \partial_x q\, \partial_x p
+(1/2) \,\sigma(q)\!\left(\partial_x p\right)^2.
\end{equation}
Once $q(x,t)$ and $p(x,t)$ have been found, one can calculate the mechanical action $\int\int dt dx \left(p\,\partial_t q - \mathcal{H}\right)$ which, after some algebra, becomes \cite{Tailleur,DG2009b,KM_var}:
\begin{eqnarray}
  s  = \frac{1}{2}\int_0^1 dt \int_{-\infty}^\infty dx \,\sigma(q) (\partial_x p)^2. \label{action0}
\end{eqnarray}
The boundary condition for $q(x,t)$ at $t=0$ is given by $n(x,t=0)$ from Eq.~\eqref{step}.  By varying $q(x,t=1)$ and minimizing $s$ under the integral constraint~(\ref{current0}), Derrida and Gerschenfeld \cite{DG2009b} obtained
the second boundary condition:
\begin{equation}
\label{p_step}
p(x,t=1) = \lambda \,\theta(x).
\end{equation}
Here $\theta(x)$ is the Heaviside step function, and
the Lagrange multiplier
$\lambda>0$ is fixed by Eq.~(\ref{current0}).  Once $s$ is
known, ${\cal P}(J,T)$ is given, up to a pre-exponent,  by $\ln {\cal P}(J,T) \simeq -\sqrt{T} \,s(\tilde{j},n_{-},n_{+})$ \cite{DG2009b,KM_var}.  Alternatively, one can evaluate the moment generating function of $J$:
\begin{equation}
  \left\langle e^{\lambda J} \right\rangle = \sum_{J\geq 0} e^{\lambda J} {\cal P}(J,T) \sim  \int_0^{\infty} d\tilde{j} e^{\sqrt{T}\left[\lambda \tilde{j}-s(\tilde{j},n_-,n_+)\right]}
  \sim  e^{\sqrt{T}\,\mu(\lambda,n_-,n_+)}, \label{GF}
\end{equation}
where
\begin{equation}\label{GF1}
 \mu(\lambda,n_-,n_+)= \max_{\tilde{j}}\, [\lambda \tilde{j}-s(\tilde{j},n_-,n_+)].
\end{equation}

For large currents, $\tilde{j} \to \infty$, an important role is played by the inviscid version of Eqs.~(\ref{q:eqfull}) and (\ref{p:eqfull}) \cite{MS2013},
obtained when neglecting the diffusion terms:
\begin{eqnarray}
\partial_t q &=& -  \partial_x \left[\sigma(q)\, \partial_x p\right],\label{q:inviscid}\\
\partial_t p &=& - \frac{1}{2}\sigma^{\prime}(q)\!\left(\partial_x p\right)^2.\label{p:inviscid}
\end{eqnarray}
Whether the inviscid equations are hyperbolic, elliptic or parabolic is determined by the
sign of $\sigma^{\prime\prime}(q)$, see Appendix A.

Now we return to the complete MFT equations~(\ref{q:eqfull}) and (\ref{p:eqfull}), and write them down for the KMP model, where $D=1$ and $\sigma(q)=2q^2$ \cite{4n2}:
\begin{eqnarray}
  \partial_t q &=& \partial_x (\partial_x q-2 q^2 v), \label{d1} \\
  \partial_t v &=& \partial_x (-\partial_{x} v-2 q v^2), \label{d2}
\end{eqnarray}
where we have differentiated Eq.~(\ref{p:eqfull}) with respect to $x$ and introduced  the momentum density gradient $v(x,t)=\partial_x p(x,t)$.
The Hamiltonian becomes $H=\int_{-\infty}^{\infty} dx\, h$, where
\begin{equation}
h = -v \partial_x q +q^2 v^2,
\label{Q010}
\end{equation}
whereas the action (\ref{action0}) becomes
\begin{eqnarray}
  s  = \int_0^1 dt \int_{-\infty}^\infty dx \,q^2 v^2. \label{action1}
\end{eqnarray}
Importantly,  $n_-$ can be set to unity without loss of generality. This is achieved by a simple canonical transformation of rescaling $q$ by $n_-$ and $p$ (or $v$) by $1/n_-$. Equations~(\ref{p_step}) and (\ref{d1})-(\ref{action1}) remain unchanged except that $\lambda$ is replaced by $\Lambda= \lambda n_{-}$, while Eqs.~(\ref{step}) and (\ref{current0}) become
\begin{equation}
\label{step1}
q(x,t=0) =
\begin{cases}
1,   & x<0,\\
n_+/n_-,  & x>0
\end{cases}
\end{equation}
and
\begin{equation}
\label{current1}
\int_0^\infty dx\, \left[q(x,t=1)-\frac{n_{+}}{n_{-}}\right] = \frac{J}{n_-\sqrt{T}} =\frac{\tilde{j}}{n_{-}}=j,
\end{equation}
respectively. Finally, the boundary condition $p(x,1)=\Lambda \,\delta(x)$ can be rewritten as
\begin{equation}\label{delta}
    v(x,t=1) =\Lambda \,\delta(x),
\end{equation}
where $\delta(x)$ is the Dirac's delta function. All this implies scaling behavior
\begin{equation}\label{scalingb}
\ln {\cal P}(J,T,n_-,n_+)\simeq
-\sqrt{T} \,s\left(\frac{J}{n_-\sqrt{T}}, \,\frac{n_+}{n_-}\right).
\end{equation}

Importantly, the $q$ and $v$ fields are both locally conserved, as it is evident from Eqs. (\ref{d1}) and (\ref{d2}).
The exact Hamiltonian~(\ref{Q010}) is only conserved globally. However, in the inviscid flow regions, where
one can neglect the diffusion terms, the reduced Hamiltonian density
$\rho=q^2 v^2$  is approximately conserved \emph{locally} \cite{MS2013}, as it evolves by the continuity equation
\begin{equation}
\partial_t \rho+\partial_x (4 \rho^{3/2})=0.
\label{Q020}
\end{equation}
These three conservation laws will play a crucial role in the following.

\section{KMP model: inviscid equations and singularity}
\label{singul}

\begin{figure}[ht]
\includegraphics[width=3.0 in,clip=]{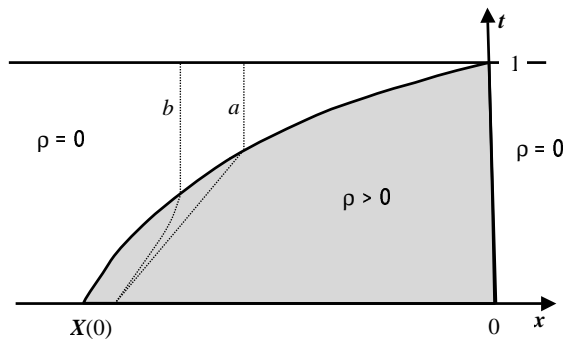}
\caption{Different solution domains of Eqs.~(\ref{Q020a}) and (\ref{Q020b}) on the $(x,t)$ plane, see the text. Letters
a and b mark characteristics of Eqs.~(\ref{Q020a}) and (\ref{Q020b}), respectively.} \label{char}
\end{figure}

That the inviscid problem for the KMP model must have a singularity at all times can be seen from
two evolution equations which follow from the inviscid equations (\ref{q:inviscid}) and (\ref{p:inviscid}).  The first of them is Eq.~(\ref{Q020}) which we rewrite as
\begin{equation}
\partial_t \rho+6 \rho^{1/2} \partial_x \rho=0,
\label{Q020a}
\end{equation}
a variant of the Hopf equation. The second one is a linear equation for $\psi=v^2 \rho^{-1/2} = v/\rho$
that one can verify directly:
\begin{equation}
\partial_t \psi+2 \rho^{1/2} \partial_x \psi=0.
\label{Q020b}
\end{equation}
The characteristics of Eq.~(\ref{Q020a}) on the $(x,t)$ plane are governed by the ordinary differential equation (ODE)
\begin{equation}\label{char1}
6 \,\sqrt{\rho} \,dt=dx .
\end{equation}
The solution of Eq.~(\ref{Q020a}) must be constant along these characteristics,
so the characteristics are straight lines. In view of the boundary condition (\ref{delta}), $\rho(x,t=1)$ is positive (and singular) at $x=0$, and is zero elsewhere.  Therefore,
there must be two distinct regions in the inviscid problem: the region of $\rho(x,t)>0$ where the characteristics of Eq.~(\ref{Q020a}) have a finite positive slope, and the region of
$\rho(x,t)=0$, where the characteristics of Eq.~(\ref{Q020a}) are parallel to the $t$-axis, see Fig.~\ref{char}.

As the characteristics of Eq.~(\ref{Q020b}) cross the (a priori unknown) boundary of the region where $\rho>0$, the function $\rho(x,t)$ (which is constant and positive along the characteristics) must have a discontinuity in every point of the boundary. Otherwise $\rho$ would be equal to zero at this boundary, and the boundary would be parallel to the $t$-axis by virtue of Eq.~(\ref{char1}). [The characteristics of Eq.~(\ref{Q020b}), described by the equation
\begin{equation}\label{char2}
2 \sqrt{\rho}\, dt=dx ,
\end{equation}
have a positive slope in the region where $\rho(x,t)>0$, so the whole solution is uniquely defined everywhere except at the discontinuity.] As we will see shortly, the discontinuity is much stronger than a shock, as both $q$ and $v$ are infinite in the inviscid limit. We will first show it by taking diffusion into account, finding uniformly translating solutions and then sending $j$ to infinity (which is equivalent to sending diffusion to zero).  In section \ref{perturb} B we
will present an additional argument by applying conservation laws to a point-like singularity in an inviscid flow. We will also show that the border line between the regions of $\rho>0$ and $\rho=0$  in Fig.~\ref{char} is actually a straight line, see Eq.~(\ref{cparam}) below.

\section{Double and single traveling pulse solutions}
\label{exactsol}

When regularized by diffusion, the singularity, encountered in the inviscid theory for the KMP model, gives way to traveling pulses of $q$ and $v$ with large but finite amplitudes. To begin with, the account of diffusion brings about three families of exact uniformly translating solutions of Eqs.~(\ref{d1}) and (\ref{d2}) that we now present. The first family describes what we call double pulses, as the solution involves finite-amplitude pulses of both $q$ and $v$. A double pulse moves on constant positive pedestals $q=q_0$ and $v=v_0$. The closely related second and third families of exact solutions describe single pulses: of $q$ and $v$, respectively, while the other field ($v$ and $q$, respectively) varies monotonically with $x$. Neither of these exact traveling pulse solutions satisfies the boundary conditions of our extreme current statistics problem. Furthermore, because of the constant pedestals these solutions would lead to an infinite  action (\ref{action1}). Still, they are instrumental in our subsequent construction of a \emph{perturbative} solution to the extreme current statistics problem, based on the small parameter $1/j$.

\begin{figure}[ht]
\includegraphics[width=2.5 in,clip=]{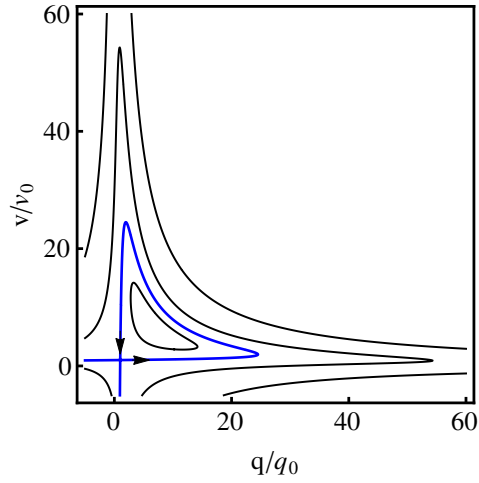}
\caption{The phase portrait of Eqs.~(\protect\ref{Q030}) and (\protect\ref{Q040}) for $R=c/(q_0v_0)=100$. The homoclinic trajectory (thick line) corresponds to a double traveling pulse solution.}  \label{phaseport}
\end{figure}

\begin{figure}[ht]
\includegraphics[width=2.5 in,clip=]{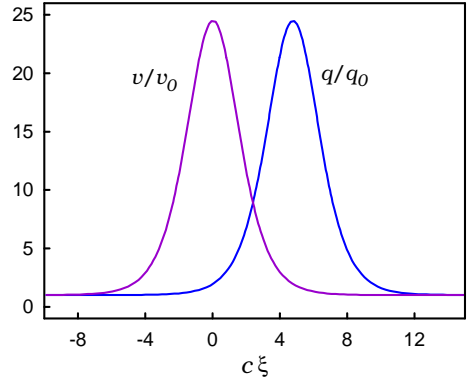}
\caption{The rescaled $q$ and $v$ profiles of the double traveling pulse solution for  $R=c/(q_0v_0)=100$.}  \label{KMP2}
\end{figure}

Consider uniformly translating solutions of Eqs.~(\ref{d1}) and~(\ref{d2}): $\{q(x,t), v(x,t)\}=\{q(\xi), v(\xi)\}$, where $\xi=x-ct$ and $c=\text{const}>0$. Plugging this ansatz into Eqs.~(\ref{d1}) and (\ref{d2}) and integrating over $\xi$, we obtain two coupled first-order ODEs, which can be written as
\begin{eqnarray}
q^{\prime} &=& -cq+2q^2v +cq_0-2q_0^2v_0, \label{Q030} \\
v^{\prime} &=& cv-2qv^2-cv_0+2q_0v_0^2, \label{Q040}
\end{eqnarray}
where $q_0=\text{const}$ and $v_0=\text{const}$. These ODEs are
Hamiltonian, $q^{\prime}=\partial_v H_0(q,v)$ and $v^{\prime}=-\partial_q H_0(q,v)$,
with the Hamiltonian
\begin{equation}
H_0(q,v)=qv (qv-c)+(c-2q_0v_0)(v_0q+q_0v).
\label{Q050}
\end{equation}
As $H_0(q,v)$ is an integral of motion,
\begin{equation}\label{constant}
    H_0(q,v)=\text{const},
\end{equation}
Eqs.~(\ref{Q030}) and (\ref{Q040}) are integrable in quadratures. A double pulse solution is a particular solution which involves a rigidly coupled pair of
solitary pulses of $q$ and $v$,
propagating with speed $c>0$ on a constant positive pedestal: $q(\xi \to \pm \infty)=q_0>0$ and $v(\xi \to \pm \infty)=v_0>0$, so that $q(\xi)>q_0$ and $v(\xi)>v_0$ everywhere.  For such solutions one obtains
\begin{equation}
H_0(q,v)=q_0v_0(c-3q_0v_0).
\label{Q060}
\end{equation}
At fixed $q_0$, $v_0$ and $c$ the value of constant in Eq.~(\ref{Q060}) corresponds to a homoclinic trajectory on the phase plane $(q,v)$, see Fig.~\ref{phaseport}. As we will see shortly, when properly rescaled,
Eqs.~(\ref{Q030}) and (\ref{Q040}) are
governed by a single parameter $R=c/(q_0 v_0)$. Homoclinic trajectories only exist for $R>6$, or $c>6 q_0 v_0$. The characteristic width of the $q$ and $v$ pulses in the double solution  scales as $1/c$.  Figure~\ref{KMP2} depicts the rescaled $q$ and $v$ profiles of a double pulse solution. As one can see, the $q$ pulse is always trailed by the $v$ pulse.

Upon
rescaling $Q=q/q_0$, $W=v/v_0$ and $z=c\xi$, Eqs.~(\ref{Q030}) and~(\ref{Q040}) become
\begin{eqnarray}
Q^{\prime}(z) &=& -Q +\frac{2Q^2W}{R} +1-\frac{2}{R}, \label{Q030a} \\
W^{\prime}(z) &=& W-\frac{2QW^2}{R}-1+\frac{2}{R}. \label{Q040a}
\end{eqnarray}
For the homoclinic trajectory the integral of motion is
\begin{equation}
QW (QW-R)+(R-2)(Q+W) =R-3.
\label{Q060a}
\end{equation}
As we will see below, a large rescaled current $j$ corresponds to a large $c$. Therefore, of most interest for us is the
regime of $R\gg 1$, when the $q$ and $v$ pulses are fast, large-amplitude and narrow. Here Eqs.~(\ref{Q030a}) and (\ref{Q040a}) can be approximated  as
\begin{eqnarray}
Q^{\prime}(z) &=& -Q +\frac{2Q^2W}{R} +1, \label{Q030b} \\
W^{\prime}(z) &=& W-\frac{2QW^2}{R}-1, \label{Q040b}
\end{eqnarray}
whereas Eq.~(\ref{Q060a}) becomes
\begin{equation}
QW (QW-R)+R(Q+W) \simeq 0.
\label{Q060b}
\end{equation}
Using Eqs.~(\ref{Q030b})-(\ref{Q060b}), we can determine the rescaled amplitudes $Q_m=q_m/q_0$ and $W_m=v_m/v_0$ of the $q$ and $v$ pulses of the double pulse solution. In the leading order
\begin{equation}\label{Qm_Wm}
    Q_m=W_m\simeq \frac{R}{4}\gg 1.
\end{equation}
When $R \to \infty$, the $q$ and $v$ pulses become infinitely high and narrow, while the distance between them tends to zero. This is the type of singularity one has to deal with in the inviscid description.

For $R\gg 1$ but finite the $q$ and $v$ pulses of the double solution are well separated: $q$ is much smaller than $q_m$ in the location of the $v$ pulse, and $v$ is much smaller than $v_m$ in the location of the $q$ pulse. To obtain the corresponding asymptotics of $q$ and $v$, we consider separately two regions: $Q\ll R$, and $W\ll R$.

\subsubsection{$v$ pulse region: $q\ll q_m$, or $Q\ll R$}
Here Eq.~(\ref{Q060b}) can be approximated as
\begin{equation}
QW (QW-R)+RW = 0,
\label{Q060e}
\end{equation}
or
\begin{equation}
W=R(Q^{-1}-Q^{-2}).
\label{Q060f}
\end{equation}
Using this relation in Eq.~(\ref{Q030b}), we obtain $Q^{\prime}(z)=Q-1$. Therefore,
\begin{equation}\label{QWaa}
    Q(z)=e^z+1,\;\;\; W(z)= \frac{R}{4}\, \cosh^{-2} \left(\frac{z}{2}\right).
\end{equation}
where we have chosen the arbitrary constant by fixing the $z$ (or $\xi$) axis so that $W(z=0)=W_m = R/4$.

\subsubsection{$q$ pulse region: $v\ll v_m$, or $W\ll R$}
Here we can approximate Eq.~(\ref{Q060b}) as
\begin{equation}
QW (QW-R)+RQ = 0,
\label{Q060c}
\end{equation}
or
\begin{equation}
Q= R(W^{-1}-W^{-2}).
\label{Q060d}
\end{equation}
This leads to $W^{\prime}(z)=-W+1$ and
\begin{equation}\label{Wb}
W(z)=e^{z_0-z}+1,
\end{equation}
where $z_0$ is a constant to be determined. Equations~(\ref{Q060d}) and~(\ref{Wb}) yield
\begin{equation}\label{Qb}
    Q(z)= \frac{R}{4} \cosh^{-2} \left(\frac{z-z_0}{2}\right).
\end{equation}

\subsubsection{Intermediate region: $q_0\ll q\ll q_m$ and  $v_0\ll v\ll v_m$}
The unknown constant $z_0$ can be determined by considering the intermediate region,
where the strong inequalities $q_0\ll q\ll q_m$ and  $v_0\ll v\ll v_m$ (or, equivalently, the strong
inequalities $1\ll Q\ll R$ and $1\ll W\ll R$) hold simultaneously. Matching the $z\gg 1$ asymptotic
of $W(z)$ from Eq.~(\ref{QWaa}) and the $z_0-z\gg 1$ asymptotic of Eq.~(\ref{Wb}), we obtain $z_0\simeq\ln R$.
The same result is obtained when matching the $z\gg 1$ asymptotic of $Q(z)$ from Eq.~(\ref{QWaa}) and the
$z_0-z\gg 1$ asymptotic of Eq.~(\ref{Qb}). In the intermediate region Eq.~(\ref{Q060b}) becomes
\begin{equation}
QW (QW-R) = 0,
\label{Q060int}
\end{equation}
which yields $QW=R$, a hyperbola.

\subsubsection{Single $q$ and $v$ pulses}

Now let us return to Eqs. (\ref{Q030}) - (\ref{Q050}) and (\ref{Q060}) and find two more families of exact traveling pulse solutions. Each of them involves a single pulse, of either $q$ or $v$, whereas the second field  is monotonic in $\xi$.  For the $v$ pulse $q_0>0$ but $v_0=0$, for the $q$ pulse $q_0=0$ but $v_0>0$.  The corresponding solutions, up to arbitrary shifts in $\xi$, are
\begin{equation}
  q(\xi)=q_0 \,(1+e^{c \xi}),\;\;\;v(\xi)=\frac{c}{4 q_0}  \cosh^{-2} \left(\frac{c \xi}{2}\right) \label{vpulse}
\end{equation}
for the $v$ pulse, and
\begin{equation}
  q(\xi) = \frac{c}{4 v_0}  \cosh^{-2} \left(\frac{c \xi}{2}\right),\;\;\; v(\xi)=v_0 \,(1+e^{-c \xi}) \label{qpulse}
\end{equation}
for the $q$ pulse. For the $v$ pulse  $v$ vanishes at $\xi \to \pm \infty$, while $q$ approaches $q_0$ at $\xi \to -\infty$ and diverges at $\xi \to \infty$. For the $q$ pulse $q$ vanishes at $\xi \to \pm \infty$, while $v$ approaches $v_0$ at $\xi \to \infty$ and diverges
at $\xi \to -\infty$. When comparing Eq.~(\ref{vpulse}) with Eq.~(\ref{QWaa}), and Eq.~(\ref{qpulse}) with Eqs.~(\ref{Wb}) and (\ref{Qb}), we observe that the pairs of equations coincide, up to a rescaling and shift in $\xi$. That is, Eqs.~(\ref{QWaa}), as well as Eqs.~ (\ref{Wb}) and (\ref{Qb}), which we obtained as $R\gg 1$ asymptotics of the exact double pulse solution, are by themselves \emph{exact} solutions of Eqs. (\ref{Q030}) and (\ref{Q040}): for $q_0=0$ or $v_0=0$, respectively. Alternatively, the exact double pulse solution can be approximately described,
at $R\gg 1$, in terms of two exact single pulse solutions, for $q$ and for $v$,  with a proper choice
of the distance between the pulses. We will extend this idea shortly in the context of a perturbative solution of the extreme current statistics problem.

\section{Perturbative solution to the extreme current statistics problem}
\label{perturb}

\subsection{General}
As we have already noticed, neither of the exact traveling pulse solutions of the KMP model -- double or single -- can satisfy the boundary conditions of the extreme current statistics problem. Furthermore, these solutions lead to an infinite action. Here we construct a perturbative solution based on the small parameter $1/j$. Up to small corrections (that we will not be interested in) the \emph{internal} solution, which fully resolves the singularity, represents two coupled single traveling pulses $q$ and $v$ \emph{with slowly varying parameters}. As in the exact double pulse solution, the $q$ pulse is leading: it collects most of the available energy on its way and brings it to the origin at $t=1$. The $v$ pulse is trailing. The distance between the pulses slowly increases with time but, at large $j$, remains much smaller than the size of the inviscid \emph{external} regions. Furthermore, although the distance between the pulses goes to zero as $j\to \infty$, the width of the individual $q$ and $v$ pulses goes down faster, so the pulses are well separated at large $j$.  Outside of the combined pulse the solution of Eqs. (\ref{d1}) and~(\ref{d2}) is large-scale: the size of the inviscid external regions scales as $j\gg 1$. Here diffusion is negligible, and we can write
\begin{eqnarray}
  \partial_t q &+& \partial_x (2 q^2 v)=0, \label{d10} \\
  \partial_t v &+& \partial_x (2 q v^2)=0. \label{d20}
\end{eqnarray}
At $j\gg 1$ the flow in the inviscid regions is quite simple, see below.   When viewed from the inviscid regions, the  coupled pulses of $q$ and $v$ can be regarded as a single effective point-like singularity
located at a point $x=X(t)$.  The point-like singularity carries a finite energy $M$, a finite momentum jump $V=p_1-p_2$ (where the subscripts $1$ and $2$ refer to ``in front of the singularity" and ``behind the singularity", respectively), and a finite value of the Hamiltonian ${\cal H}$. The formal definitions of these quantities are
\begin{equation}\label{PM060}
    \{M(t),\,V(t),\,{\cal H}(t)\} = \int_{X(t)-\varepsilon}^{X(t)+\varepsilon} \{q,\,v,\,h\}\, dx,
\end{equation}
where $\varepsilon$ can be fixed at an arbitrary small value as $j\to\infty$. The instantaneous pulse speed $c(t)=\dot{X}(t)$ is allowed to slowly vary in time (the upper dots denote the time derivatives).  The variations with time of $M(t)$, $V(t)$ and ${\cal H}(t)$ are also very slow:  they are small during the time it takes the combined pulse to pass the distance between the $q$ and $v$ pulses. This and other assumptions of the perturbation theory break down at times very close to $t=0$ and $t=1$, but these regions give only a small contribution to the action that can be neglected. We  employ conservation laws to properly match the point-like singularity with the inviscid external regions. This procedure yields evolution equations for the slowly varying $M(t)$, $V(t)$ and ${\cal H}(t)$. We express $M$, $V$ and ${\cal H}$ via the speed of the combined pulse $c=\dot{X}$ and the amplitudes $q_m$ and $v_m$ of the $q$
and $v$ pulses, respectively,  by considering the internal region (where diffusion is fully accounted for) and using our approximate solutions for the $q$ and $v$ pulses.  Solving the evolution equations with a proper boundary condition in time,  we determine all attributes of the approximate perturbative solution. This enables us to evaluate the action (\ref{action1}) and $\ln {\cal P}$.

\subsection{Point-like singularity in an inviscid flow}

Consider an effective point-like singularity propagating in an inviscid ``KMP medium". At time $t$  the singularity is located at $x=X(t)$. Correspondingly, at $0<t<1$, there are four external regions which can be regarded as inviscid: $x<X(0)$, $X(0)<x<X(t)$, $X(t)<x<0$ and $x>0$.
The hyperbolic character of the inviscid problem (\ref{d10}) and (\ref{d20})
implies that $v=0$ for $x<X(t)$ and $x>0$.  Then Eq.~(\ref{d10}) yields $\partial_t q =0$ in these regions. For $X(0)<x<X(t)$ this yields $q(x,t)=q_{<}(x)$, where $q_{<}(x)$ is an unknown time-independent function. For $x<X(0)$ and $x>0$  the energy density remains the same as at $t=0$, that is $q(x,t)=1$ and $q(x,t)=n_+/n_-$, respectively, so we do not need to deal with these two regions.
That the region $x>0$ does not contribute to the action immediately implies that the function $s$ in Eq.~(\ref{scalingb}) is independent  of the argument $n_+/n_-$, and so $\ln {\cal P}$ is independent of $n_+$.

The region $X(t)<x<0$ is governed by the inviscid equations~(\ref{d10}) and (\ref{d20}). To remind the reader, we
denote by subscripts 1 and 2 the values of the $q$ and $v$ fields in front of and behind the point-like singularity, respectively. There are four such quantities: $q_1(t)$ and $v_1(t)$ in front of the singularity, and $q_2(t)$ and $v_2(t)$ behind it.
As we have already observed,  $v_2(t)=0$. From the viewpoint of the internal solution, the quantities
$q_1(t)$ and $v_1(t)$ correspond to the values of the fields at $\xi =\infty$, whereas the quantities $q_2(t)$ and $v_2(t)=0$ correspond to the values at $\xi =-\infty$. Here $\xi$ is defined, up to an additive constant, as $\xi = x-X(t) = x-\int_0^t dt^{\prime}\, c(t^{\prime})$.

Differentiating $M, V$ and ${\cal H}$ from Eqs.~(\ref{PM060}) with respect to time and using Eqs.~(\ref{d1})-(\ref{Q010}) and the equality $v_2=0$, we obtain
\begin{eqnarray}
\dot{M} &=& (q_1-q_2) \dot{X}-2 q_1^2v_1,
\label{PM110}\\
\dot{V} &=& v_1 \dot{X} -2 q_1 v_1^2,
\label{PM112}\\
\dot{{\cal H}} &=& q_1^2v_1^2 \left(\dot{X} -4 q_1 v_1\right).
\label{PM114}
\end{eqnarray}
Note that the dominating contribution to the right hand side of  Eq.~(\ref{PM114}) comes from the inviscid region
where $h$ is conserved locally, see Eq.~(\ref{Q020}).

Equations~(\ref{PM110})-(\ref{PM114}) imply, independently from the regularization procedure of the previous subsection, that the effective point-like singularity of the inviscid description is stronger than the shock discontinuities of conventional hydrodynamics \cite{LLfluidmech}.
Indeed, let us assume the opposite for a moment: that $M=V={\cal H}=\dot{M}=\dot{V}=\dot{{\cal H}}=0$.  Then Eqs.~(\ref{PM110})-(\ref{PM114}) yield, at $v_1\neq 0$,
 \begin{equation*}
\dot{X}=\frac{2 q_1^2 v_1}{q_1-q_2},\;\;\;\dot{X}=2 q_1 v_1,\;\;\text{and}\;\;\dot{X}=4 q_1 v_1.
 \end{equation*}
At $q_1\neq 0$ these equations are obviously contradictory.

Equations (\ref{PM110})-(\ref{PM114}) are only valid when the distance between the $q$ and $v$ pulses is much smaller than
the characteristic sizes of the inviscid regions.  As can be checked a posteriori, this strong inequality demands $j\gg 1$. In this case, however, the amplitude of the $v$ pulse, $v_m$, is much larger than $v_1$,  whereas
the amplitude of the $q$ pulse, $q_m$, is much larger than $q_1$.  To avoid excess of accuracy, we must simplify Eqs.~(\ref{d10}) and (\ref{d20}) in the region $X(t)<x<0$, and also simplify Eqs.~(\ref{PM110})-(\ref{PM114}). In the region $X(t)<x<0$ we can neglect the term quadratic in $v$ in Eq.~(\ref{d20}). This yields $\partial_t v = 0$. In Eq.~(\ref{d10}) the term including $v$ is subleading and can also be neglected. As a result, we obtain
\begin{equation}
q(x,t) = 1,\;\;v(x,t)=v_{>} (x)\;\;\text{at}\;\;X(t)<x<0, \label{PM030a}
\end{equation}
where $v_{>}(x)$ is a time-independent function. The equations for $x<X(t)$ do not change:
\begin{equation}
  q(x,t) = q_{<}(x),\;\;v(x,t)=0\;\;\text{at}\;\;x<X(t). \label{PM010a}\\
\end{equation}
Neglecting all but one terms including $v_1$ in Eqs.~(\ref{PM110})-(\ref{PM114}), we arrive at simplified equations for the parameters of the effective
singularity:
\begin{eqnarray}
\dot{M} &=& (1-q_2) \dot{X},
\label{PM110a}\\
\dot{V} &=& v_1 \dot{X},
\label{PM112a}\\
\dot{{\cal H}} &=&0,
\label{PM114a}
\end{eqnarray}
Finally, one has $v_1(t)=v_>[x=X(t)+0]$ and $q_2(t)=q_<[x=X(t)-0]$.
To have a complete description, we need to consider the internal solution and express the quantities $M$, $V$ and $\cal H$ through its parameters.

\subsection{Slowly evolving coupled $q$ and $v$ pulses}

\begin{figure}
\includegraphics[width=3.5in,clip=]{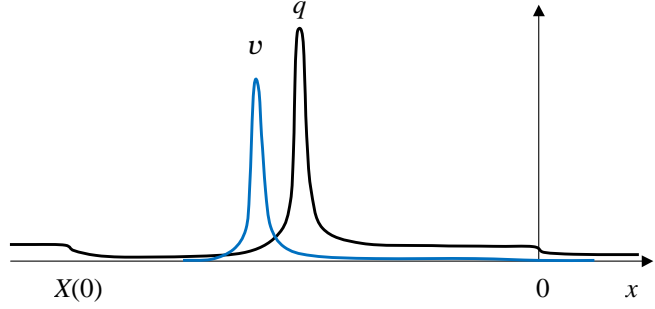}
\caption{A sketch of the regularized singularity -- the coupled $q$ and $v$ pulses with slowly varying parameters -- propagating in an inviscid ``KMP medium".}
\label{sketch}
\end{figure}

The coupled pulse solution $\{q_{\text{pulse}}(\xi), v_{\text{pulse}}(\xi)\}$ consists of a large-amplitude $q$-pulse trailed by a large-amplitude $v$-pulse, see Fig. \ref{sketch}.  In the light of results of the previous subsection, the $q$ pulse must obey the boundary conditions $q(\xi \to \infty)=1$ and $v(\xi \to \infty)=v_1$, whereas the $v$ pulse must obey  $q(\xi \to -\infty)=q_2$ and $v(\xi \to -\infty)=0$. The quantities $v_1$ and $q_2$ slowly vary with time. Therefore, the parameters of the coupled pulse solution also slowly vary in time.
In view of Eqs.~(\ref{vpulse}) the solution takes the form
\begin{eqnarray}
  q_{<}(\xi)&=&q_2 (1+e^{c \xi}), \label{vpulse1a}\\
  v_{<}(\xi) &=& v_m \cosh^{-2} \,\frac{c \xi}{2}, \label{vpulse1b} \\
  v_m &=&\frac{c}{4 q_2} \gg 1 \gg v_1, \label{vpulse1c}
\end{eqnarray}
where $c=c(t)$, $q_2=q_2(t)$ and $v_m=v_m(t)$, and we have fixed $\xi$ so that $\xi=0$ is at the maximum of $v$. The $v$-pulse region obeys the strong inequality $\xi_0-\xi\gg 1/c$, where  $\xi_0$ will be introduced shortly.

In the region of the $q$ pulse, $\xi \gg 1/c$,  the solution is described by the equations
\begin{eqnarray}
  q_{>}(\xi)&=&q_m \cosh^{-2} \,\frac{c(\xi- \xi_0)}{2}, \label{qpulse1a} \\
  v_{>}(\xi)&=&v_1 \left[1+e^{c(\xi_0-\xi)}\right], \label{qpulse1b} \\
  q_m &=&\frac{c}{4 v_1} \gg 1 \gg q_2, \label{qpulse1c}
\end{eqnarray}
with slowly time-dependent $v_1(t)$ and $q_m(t)$. Note that $q_{>}(\xi)$ tends to zero, and not to 1, as $\xi \to \infty$. This mismatch can be taken care of in the next order of perturbation theory which would account for small corrections to the pulse solutions (\ref{vpulse1a}) and (\ref{vpulse1b}), and (\ref{qpulse1a}) and (\ref{qpulse1b}). We are not interested in these small corrections here, as they only give a subleading contribution to the action, which we ignore.

The criteria for slow evolution of the coupled pulses are
\begin{equation}\label{crit}
 \frac{\dot{v}_m}{v_m}\ll c^2,\quad \frac{\dot{q}_m}{q_m}\ll c^2, \quad
\frac{\dot{c}}{c}\ll c^2.
\end{equation}
We have ignoreed in these strong inequalities logarithmic factors in $c$, see below, by using the characteristic width of the $q$ and $v$ pulses, ${\cal O}(1/c)$, instead of the distance between them. As the time scale of the parameter variation is of the order of 1, each of the three criteria in Eq.~(\ref{crit})
boils down to the strong inequality $c\gg 1$.  (As we will see later, one can parameterize the solution
by either one of the three parameters: $j\gg 1$, $\Lambda\gg 1$,  or $c\gg 1$. Each of them can be expressed via the other.)

Now we calculate the distance $\xi_0=\xi_0(c,q_m,v_m)$ between the $q$ and $v$ pulses. This can be done by
matching the solutions (\ref{vpulse1a}) and (\ref{qpulse1a}) or, alternatively, (\ref{vpulse1b}) and (\ref{qpulse1b}), in the joint region of their validity,
$\min\{\xi,\xi_0-\xi\}\gg 1/c$, where $c$ is assumed to be much greater than one.   As we neglect small corrections to the pulse solution,
we can only perform the matching in the leading order, and
the result is
\begin{equation}
\xi_0(t) =\frac{1}{c}\ln\frac{16q_mv_m}{c}.
\label{PM200}
\end{equation}
As the argument of logarithm is very large [by virtue of the strong inequalities in Eq.~(\ref{vpulse1c}) or
Eq.~(\ref{qpulse1c})], $\xi_0$ is much larger than the characteristic width of the individual $q$ and $v$ pulses, ${\cal O}(1/c)$, so the individual pulses are well separated.  The whole combined pulse becomes point-like as $c$ goes to infinity.

Now we use Eqs. (\ref{vpulse1a})-(\ref{qpulse1c}) and (\ref{PM200}) to evaluate $M$, $V$ and $\cal H$ from Eq.~(\ref{PM060}). In terms of the internal solution, Eq.~(\ref{PM060}) can be rewritten as
\begin{equation}\label{three}
    \{M(t),\,V(t),\,{\cal H}(t)\} = \int_{-\infty}^{\infty} \{q(\xi),\,v(\xi),\,h(\xi)\}\, d\xi.
\end{equation}
The integrations are performed in Appendix B.  As expected on intuitive grounds, $M$ is mostly determined by the $q$ pulse, and $V$ is mostly determined by the $v$ pulse. The result for $\cal H$ is less intuitive, as the contribution only comes from the $q$ pulse. The results are:
\begin{equation}\label{MVH}
    M=\frac{4q_m}{c}=\frac{1}{v_1},\;\;\;V=\frac{4v_m}{c}=\frac{1}{q_2}, \;\;\text{and} \;\;{\cal H}=c.
\end{equation}
We can also calculate the contribution of the double pulse to the action (\ref{action1}). As can be checked a posteriori, the contribution to the action from the inviscid region $X(t)<x<0$ is small, so we can write
\begin{eqnarray}
  s &=& \int_0^1 dt\, \dot{s}(t),\;\;\text{where} \label{svssdot}\\
  \dot{s} (t)&=& \int_{-\infty}^{\infty} d\xi\,q^2_{\text{pulse}}(\xi) \, v^2_{\text{pulse}}(\xi). \label{sdot}
\end{eqnarray}
The integration in Eq.~(\ref{sdot}) (see Appendix B) yields
\begin{equation}
\dot{s}(t) =c^2\,\xi_0 =c\, \ln\frac{16q_mv_m}{c}= c\, \ln (cMV),
\label{PM234}
\end{equation}
where we have used Eqs.~(\ref{PM200}) and~(\ref{MVH}). As one can see, the action accumulation rate (\ref{PM234})
is mostly determined by the pulse speed $c$, as it depends on $M$ and $V$ only logarithmically. As a result,
transport of the same (very large) integrated current, during the same time, by two or more narrow pulses would require a greater action and is therefore much less probable.

\subsection{Solving the evolution equations}
Using Eqs.~(\ref{MVH}) and the equality $\dot{X}=c$, we can rewrite Eqs.~(\ref{PM110a})-(\ref{PM114a}) as three
autonomous ODEs:
\begin{eqnarray}
  \dot{M} &=& c\,\left(1-\frac{1}{V} \right), \label{PM262}\\
  \dot{V} &=& \frac{c}{M}, \label{PM264}\\
  \dot{X}&=& c, \label{dotX}
\end{eqnarray}
where $c=\text{const}$: the effective point-like singularity moves with a constant speed. The general solution of Eqs.~(\ref{PM262})-(\ref{dotX}) depends on $c$ and on three arbitrary constants which, in general, are functions of $c$. If $j$ is specified,  we need to find $c$ and the three arbitrary constants in order to have a unique solution. Therefore,  we need four conditions. (Actually, we will need five, as two of the conditions include the Lagrange multiplier $\Lambda$ which itself needs to be expressed via $j$.) As a first condition, we demand that $X(t=1)=0$ so that all the energy accumulated by the effective point-like singularity is brought to the origin at $t=1$. Then Eq.~(\ref{dotX}) yields
$X(t)=c(t-1)$. In particular, $X(0)=-c$.  Now, Eqs.~(\ref{PM262}) and (\ref{PM264}) have an integral of motion, and we obtain
\begin{equation}
M(t)=\frac{e^{V(t)}}{k_1(c)\, V(t)}\, ,
\label{PM270}
\end{equation}
where the integration constant $k_1(c)$ is a function of $c$. Plugging Eq.~(\ref{PM270}) into Eq.~(\ref{PM264}) and integrating, we obtain
\begin{equation}
\text{Ei}(V)=k_1(c) \,ct  +k_2(c)\, ,
\label{PM280}
\end{equation}
where $\text{Ei}(V)$ is the exponential integral function \cite{Wolfram}:
\begin{equation}
\text{Ei} (V)=-\int\limits_{-V}^{\infty} \frac{du\,e^{-u}}{u} \underset{V\gg 1}= e^V \left(\frac{1}{V}+\frac{1}{V^2}+\frac{2}{V^3}+\dots \right),
\label{Ei}
\end{equation}
and $k_2(c)$ is another integration constant. The second condition is obtained by integrating the boundary condition (\ref{delta}) over $x$ leading to $V(t=1)=\Lambda$.
Then Eq.~(\ref{PM280}) yields
\begin{equation}\label{delta2}
  \text{Ei}(\Lambda) =c k_1(c)+k_2(c).
\end{equation}
The third condition is obtained by using Eq.~(\ref{PM270}) at $t=1$:
\begin{equation}\label{J1}
  j=\frac{e^{\Lambda}}{k_1(c) \Lambda},
\end{equation}
where we have identified the rescaled integrated current $j$ with the total energy $M(t=1)$ brought by the point-like singularity to the origin at $t=1$. The fourth condition should follow from the initial condition $q(x,t=0)=1$, see
Eq.~(\ref{step1}). Na\"{\i}vely, one would translate this condition to the equality $M(t=0)=0$. This cannot be done, however, as $M=0$ is a singular point of Eq.~(\ref{PM264}). Therefore, we should proceed with care, and first identify the narrow boundary layer $t\ll 1$ where our perturbation theory, including Eqs.~(\ref{PM262})-(\ref{dotX}), breaks down.

The boundary layer solution must describe the formation of
coupled $q$ and $p$ pulses, starting from the flat profile $q=1$. Boundary-layer-type effects at small times (although not involving formation of pulses) are also observed for the non-interacting  random walkers (where the whole problem is exactly soluble) \cite{DG2009b,BMunpubl}, and for the SSEP \cite{MS2013,MSV2013}. A defining feature of the boundary layer solution is that all terms of Eqs.~(\ref{d1}) and (\ref{d2}) are comparable with each other. As $q \sim 1$ at these short times, we obtain, from each of Eqs.~(\ref{d1}) and (\ref{d2}), the following order-of-magnitude estimates:
\begin{equation}\label{est1}
    \frac{1}{\tau}\sim \frac{1}{\ell^2} \sim \frac{v_0}{\ell},
\end{equation}
where $\tau$ is the characteristic boundary layer width (or rather duration), $\ell$ is the spatial scale of $q$ and $v$ profiles,
and $v_0$ is the magnitude of $v$.  On the other hand, we can extend, at the level of order-of-magnitude estimates,
Eqs.~(\ref{MVH}) to these short times. As $q_2 \sim q_m \sim 1$, we obtain
\begin{equation}\label{est2}
   v_m\sim v_1 \sim c,\;\;\;M\sim \frac{1}{c},\;\;\;V\sim 1.
\end{equation}
As one can see,  $M$ is still very small at these early times, whereas $V$ is already significant. Identifying $v_m\sim v_1\sim c$ with $v_0$ from (\ref{est1}), we obtain
\begin{equation}\label{est3}
\tau \sim \frac{1}{c^2},\;\;\;\ell \sim \frac{1}{c},\;\;\;v_0\sim c.
\end{equation}
Notice that $\tau$ is comparable to the time it takes the newly born pulse to travel the distance of the order of its width.
The estimates (\ref{est3}) suffice to show that the action coming from the boundary layer region is negligibly small.
Indeed, using Eq.~(\ref{action1}), we can estimate $s\sim v_0^2 \ell \tau \sim 1/c \ll 1$.  This contribution is so small that it is beyond the MFT approximation.

Let us make one more order-of-magnitude estimate, by using Eqs.~(\ref{PM270}) and (\ref{est2}) at $t=\tau$.
This yields $k_1(c)\sim c$, that is the function $k_1(c)$ is large. In its turn, using Eqs.~(\ref{PM280}) and (\ref{est2}) at $t=\tau$, we obtain $k_2(c)\sim 1$ or less. Therefore, the integration constant $k_2(c)$, which appears in Eqs.~(\ref{PM280}) and (\ref{delta2}), can be neglected at $t\gg \tau$.

Now we evaluate the action from Eqs.~(\ref{svssdot}) and (\ref{PM234}), determine the unknown function $k_1(c)$, and
obtain relations between $c$, $\Lambda$ and $j$. Plugging Eq.~(\ref{PM270}) in Eq. ~(\ref{PM234}), we obtain

\begin{equation}
\dot{s}=c\ln (cMV)=c V + c \ln\frac{c}{k_1(c)}.
\label{A020}
\end{equation}
As a result,
\begin{equation}
s=c\int\limits_0^1 V \,dt+ c \ln\frac{c}{k_1(c)}.
\label{A030}
\end{equation}
The integral entering Eq.~(\ref{A030}) can be evaluated using Eqs.~(\ref{PM264}) and (\ref{PM270}):
\begin{equation}
  \int\limits_0^1 V\, dt= \int\limits_{{\cal O}(1)}^{\Lambda} dV\, \frac{dt}{d V} V = \int\limits_{{\cal O}(1)}^{\Lambda} dV\, \frac{V}{\dot{V}} = \frac{1}{c}\int\limits_{{\cal O}(1)}^{\Lambda} dV\,M V
 = \frac{1}{c k_1(c)}\int\limits_{{\cal O}(1)}^{\Lambda} dV\,e^V \simeq \frac{e^{\Lambda}}{c k_1(c)},
\end{equation}
where we have neglected a small contribution from the lower bound of integration.  As a result,
\begin{equation}
s=\frac{e^\Lambda}{k_1(c)}+c \ln\frac{c}{k_1(c)}.
\label{A040}
\end{equation}
As the parameter $\Lambda$ is a Lagrange multiplier \cite{DG2009b}, we can use the condition
[see Eq.~(\ref{GF1})]
\begin{equation}\label{dsdj}
\frac{ds}{dj}=\Lambda
\end{equation}
as the fifth (and last) condition. It is assumed in Eq.~(\ref{dsdj}) that $c$ is expressed via $\Lambda$ and $j$ with the help of Eq.~(\ref{J1}). Now we need to find $k_1(c)$ from Eqs.~(\ref{delta2}), (\ref{J1}),
(\ref{A040}) and (\ref{dsdj}). It is convenient to introduce a new unknown function $\kappa(c)$ so that  $k_1(c)= c \kappa(c)$. After some algebra (see Appendix C) we obtain $\kappa(c)=e^2$, independent of $c$, as expected from our order-of-magnitude
estimates above. Notice that using Eq.~(\ref{dsdj}) is equivalent to minimizing the action (\ref{A040}) with respect to $k_1(c) =\kappa c$ under the constraint $j=\text{const}$.

\subsection{Results}

Plugging $k_1(c)=e^2 c$ in all relevant equations, we obtain
\begin{equation}\label{cparam}
    X(t)=c(t-1),\;\;\mbox{Ei}(V)=e^{2}\, c^2\,t\gg 1,\;\;M=\frac{e^{V-2}}{c V}.
\end{equation}
These three equations provide a complete description of the dynamics of the effective point-like singularity
as parameterized by the value of its constant speed $c\gg 1$. The quantities $v_1=1/M$ and $q_2=1/V$
follow from Eqs.~(\ref{cparam}), whereas the large-scale external fields $v_>(x)$ and $q_<(x)$ are the following:
\begin{equation}\label{HDfields}
 v_>(x)=\frac{1}{M(x/c+1)},\;\;\; q_<(x)=\frac{1}{V(x/c+1)}.
\end{equation}
The traveling pulse speed $c$ can be expressed via $\Lambda$:
\begin{equation}
c=\frac{\sqrt{\text{Ei}\left(\Lambda\right)}}{e} = e^{\frac{\Lambda}{2}-1} \left(\frac{1}{\Lambda^{1/2}}
+\frac{1}{2\Lambda^{3/2}}+\frac{7}{8\Lambda^{5/2}}+\dots \right).
\label{A090}
\end{equation}
The rescaled current $j$ and action $s$ can be also expressed via $\Lambda$:
\begin{eqnarray}
j(\Lambda) &=& \frac{e^{\Lambda-1}}{\Lambda\sqrt{\text{Ei}\left(\Lambda\right)}},
\label{A100}\\
s(\Lambda) &=& \frac{e^{\Lambda-1}}
{\sqrt{\text{Ei}\left(\Lambda\right)}}-\frac{2}{e}\,\sqrt{\text{Ei}\left(\Lambda\right)},\;\;\Lambda \gg 1.
\label{A110}
\end{eqnarray}
For convenience, we also present the first three terms of their asymptotic expansions:
\begin{eqnarray}
j(\Lambda)  &=& e^{\frac{\Lambda}{2}-1} \left(\frac{1}{\Lambda^{1/2}}-\frac{1}{2\Lambda^{3/2}}-\frac{5}{8\Lambda^{5/2}}+\dots \right),
\label{A100a}\\
s(\Lambda)  &=& e^{\frac{\Lambda}{2}-1} \left(\Lambda^{1/2}-\frac{5}{2\Lambda^{1/2}}-\frac{13}{8\Lambda^{3/2}}+\dots \right).
\label{A110a}
\end{eqnarray}

We can also calculate the large deviation function $\mu$, see Eq.~(\ref{GF1}). Restoring the original
Lagrange multiplier $\lambda$, as it appears in Eq.~(\ref{p_step}),
we obtain
\begin{equation}
  \mu (\lambda,n_{-}) = n_{-} \lambda \,j(n_{-}\lambda)
-s(n_{-}\lambda) = \frac{2}{e} \sqrt{\text{Ei} (n_{-}\lambda)},
\label{A160}
\end{equation}
independent of $n_{+}$.

Equations~(\ref{cparam}), (\ref{A090}), (\ref{A100}), (\ref{A110}) and (\ref{A160}) represent the main results of this work.
We believe that their relative error is not greater than ${\cal O}(1/j)$ as $j\to \infty$.

\begin{figure}[ht]
\includegraphics[width=2.5in,clip=]{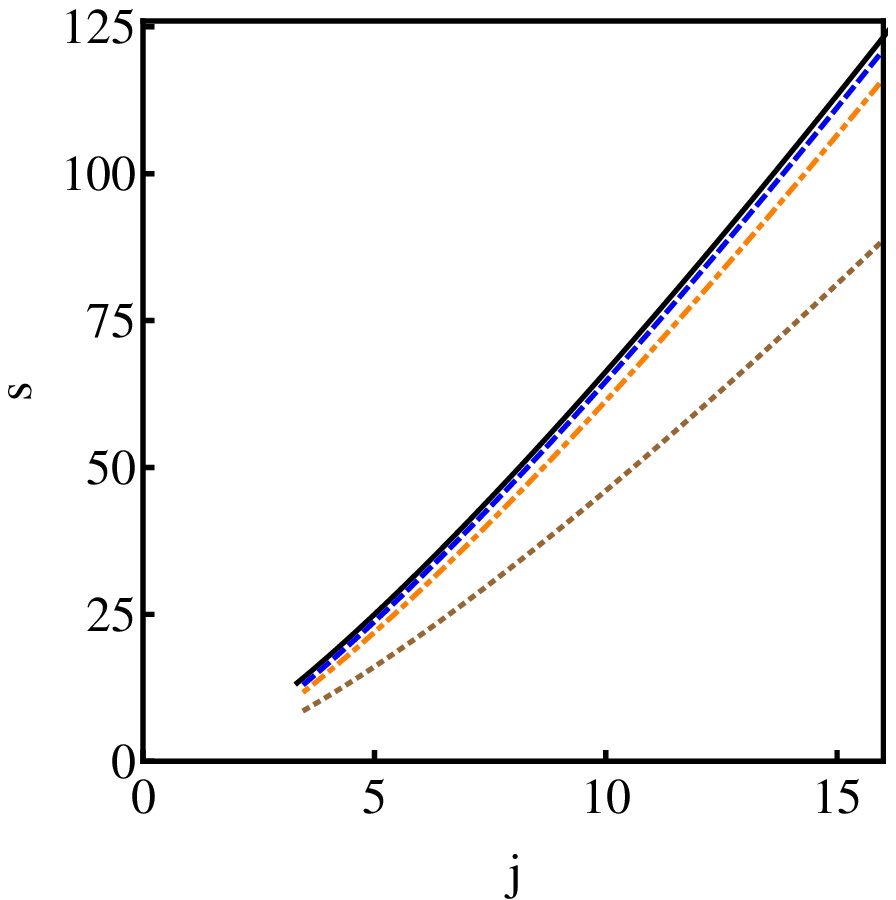}
\caption{The dependence of the rescaled action $s$ on the rescaled current $j$ as predicted by Eqs.~(\ref{A100}) and (\ref{A110}) (the solid line),
and by its one-, two- and three-term approximations from Eq.~(\ref{A170B}) (the dotted, dash-dotted and dashed lines, respectively).} \label{s(j)}
\end{figure}

Equations~(\ref{A100}) and (\ref{A110}) determine, in a parametric form, the dependence of $s(j)$ for $j\gg 1$, or $\Lambda\gg 1$.  The explicit form of this dependence can be obtained by using Eqs.~(\ref{A100a}) and (\ref{A110a}):
\begin{equation}
s=j\left[2 \ln j+\ln (2 \ln j)+\frac{\ln(2 \ln j)}{2 \ln j}+\dots\right]\, .
\label{A170B}
\end{equation}
This yields the announced asymptotic expansion (\ref{KMPexpansion}) for $\ln {\cal P}(J\to \infty,T)$.
Figure \ref{s(j)} depicts $s(j)$ from Eqs.~(\ref{A100}) and (\ref{A110})
and its one-, two- and three-term approximations from Eq.~(\ref{A170B}).

\begin{figure}[ht]
\includegraphics[width=2.2in,clip=]{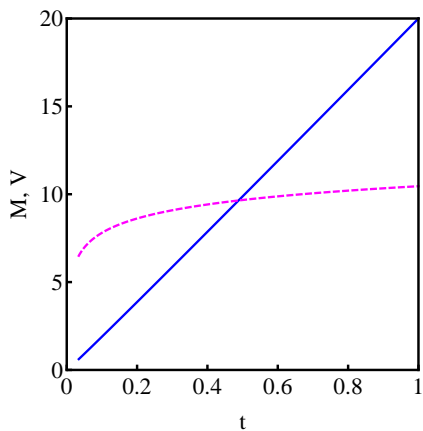}
\caption{The time dependence of the energy content $M$ (the solid line)
and the momentum jump $V=p_1-p_2$ (the dashed line) of the effective point-like singularity, as  predicted by Eq.~(\ref{cparam}) with $j=20$ (which corresponds to $c=22.47\dots$ and $\Lambda=9.939\dots$).}  \label{MV}
\end{figure}

The optimal path of the system -- the time history which dominates the contribution to the probability of observing a specified extreme current -- is mostly encoded in the time evolution of the quantities $X(t)=c(t-1)$, $M(t)$ and $V(t)$. The quantities $M(t)$ and $V(t)$ are depicted in Fig.~\ref{MV} for $j=20$. As one can see, $M(t)$ grows almost linearly in time, and much faster than $V(t)$. This can be explained by the fact that, at large $V$, $\text{Ei}(V)\simeq V^{-1} e^V$, see Eq.~(\ref{Ei}).  Then Eq.~(\ref{cparam}) predicts  simply $M(t)\simeq ct$. More accurately, the asymptotic expansions of $M(t)$ starts with
\begin{equation}
    M(t) = c t \left[1-\frac{\ln \ln (e^2 c^2 t)}{\ln (e^2 c^2 t)} + \dots \right],\;\;\;e^2 c^2 t \gg 1.
\label{Masymp}
\end{equation}
In contrast to $M(t)$,  $V$ grows with time only logarithmically:
\begin{equation}
V(t)=\ln(e^2 c^2 t)+\ln \ln(e^2 c^2 t)+\dots,\;\;\; e^2 c^2 t \gg 1.
\label{Vasymp}
\end{equation}
For most of the time $V(t)$ is close to $V(1)=\Lambda$.   [The simple results $M(t)\simeq ct$ and $V(t)\simeq \text{const} = \Lambda$ follow already from Eqs.~(\ref{PM110a}) and (\ref{PM112a}) if we neglect the small terms including $q_2$ and $v_1$.] Notice that at early times  $V(t)$ is much greater than $M(t)$: a significant buildup of fluctuations is needed already at early times to initiate the highly unusual mode of energy transfer.

\begin{figure}[ht]
\includegraphics[width=2.5in,clip=]{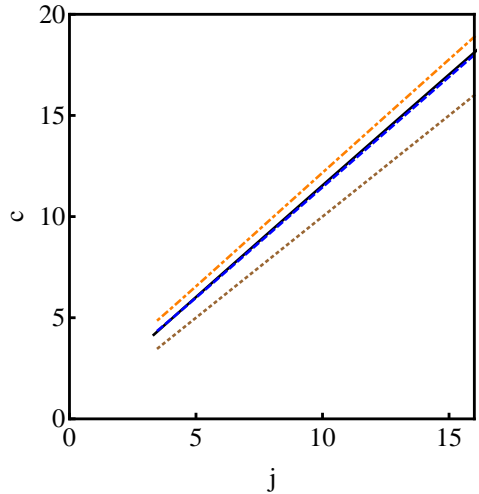}
\caption{The dependence of the pulse speed $c$ on the rescaled current $j$ as predicted by Eqs.~(\ref{A090}) and (\ref{A100}) (the solid line),
and by its one-, two- and three-term approximations from Eq.~(\ref{cvsj}) (the dotted, dash-dotted and dashed lines, respectively).} \label{c(j)}
\end{figure}

For completeness, we also present the explicit dependence of $c$ on $j$, as determined by Eqs.~(\ref{A090}) and
(\ref{A100}):
\begin{equation}
c=j\left[1+\frac{1}{2 \ln j}-\frac{\ln(2 \ln j)}{(2 \ln j)^2}+\dots\right]\, .
\label{cvsj}
\end{equation}
Figure \ref{c(j)} shows $c(j)$ from Eqs. (\ref{A090}) and
(\ref{A100}) and its one-, two- and three-term approximations from Eq. (\ref{cvsj}).

\section{Discussion}
\label{discus}

In our previous work \cite{MS2013} we uncovered two major classes of diffusive lattice gases with respect to the extreme current statistics when starting from a deterministic step-like density (or energy) profile. These two classes are determined by the sign of $\sigma^{\prime\prime}(n)$.  For gases of the elliptic class, $\sigma^{\prime\prime}(n)<0$, the extreme current statistics exhibits a universal decay,
$\ln {\cal P}(J\to \infty, T) \sim - J^3/T$ \cite{DG2009b,MS2013,MSV2013}. This decay can be tracked down to the fact that, at sufficiently large currents, one can neglect the diffusion contribution to the current compared with the contribution coming from noise. The example of KMP model, considered here, shows that for gases of the hyperbolic class, $\sigma^{\prime\prime}(n)>0$, the situation is more complicated.  Here diffusion remains relevant no matter how large the fluctuating current is.
As we have seen, the optimal (most probable) time history of the KMP model, which dominates the extreme current statistics, involves a regularized singularity: a pair of propagating large-amplitude solitary pulses of $q$ and $v$, the parameters of which slowly vary in time.  The coupled pulses are closely related to exact traveling pulse solutions of the MFT equations for $q$ and $v$, where diffusion and (the optimal realization of) noise exactly balance each other. Importantly, the dominant contribution to $\ln {\cal P}(J\to \infty, T)$ from Eq.~(\ref{KMPexpansion}) comes from the regularized singularity itself, rather than from large ``inviscid" regions. This behavior is very different from the one observed for the SSEP which belongs to the elliptic class \cite{MS2013}. Notice also that Eq.~(\ref{KMPexpansion}) for $\ln {\cal P}(J\to \infty, T)$ differs from the  corresponding result for the KMP model in the annealed setting \cite{DG2009b,PKprivate}.

By analogy with the KMP model, we expect that the extreme current statistics of other diffusive lattice gas models, belonging to the hyperbolic class, is determined by whether exact traveling pulse solutions exist in the model or not. One family of models where we can address this issue has $D(q)=1$ and
$\sigma(q)=2 q^{\alpha}$, where $\alpha \geq 1$. This family includes the non-interacting random walkers, $\alpha=1$, and the KMP model, $\alpha=2$, as particular cases.
One can show that exact traveling pulse solutions exist in these models for all $\alpha>1$ (note that $0<\alpha<1$ corresponds to the elliptic class where diffusion can be neglected altogether at large currents). This opens the way to extensions of our perturbation theory to different $\alpha>1$ and suggests a kind of universality. We do not expect, however, that the extreme current statistics~(\ref{KMPexpansion}), or its leading term, are universal (even for this $\alpha$-model).

\section*{Acknowledgments}
We thank P.L. Krapivsky for useful discussions. B.M.
was supported by the US-Israel Binational Science Foundation (Grant No. 2012145).
P.V.S. was supported by the Russian Foundation for Basic Research, grant No 13-01-00314.

\appendix

\section{Classification of inviscid MFT equations for diffusive lattice gases}
\label{aa}

Introducing $v(x,t)=\partial_x p$, we can rewrite the inviscid MFT equations (\ref{q:inviscid}) and  (\ref{p:inviscid}) as
\begin{eqnarray}
  \partial_t q +\partial_x\left[\sigma(q) v \right]&=&0, \label{A} \\
  \partial_t v +\frac{1}{2} \partial_x\left[\sigma^{\prime}(q)v^2\right]&=&0. \label{B}
\end{eqnarray}
One way of classifying this set of equation is by performing the hodograph transformation \cite{LLfluidmech}, that is by going over from $q(x,t)$ and $v(x,t)$ to $t(q,v)$ and $x(q,v)$. Equations~(\ref{A}) and (\ref{B}) become
\begin{eqnarray}
  \partial_v x&=&\sigma^{\prime}(q) v \partial_v t-\sigma(q) \partial_q t, \label{Ahod} \\
  \partial_q x&=& -\frac{1}{2} \sigma^{\prime\prime}(q) v^2 \partial_v t+\sigma^{\prime}(q) v \partial_q t. \label{Bhod}
\end{eqnarray}
Differentiating Eq.~(\ref{Ahod}) with respect to $q$ and  Eq.~(\ref{Bhod}) with respect to $v$, we obtain a linear second-order PDE for the function $t(q,v)$:
\begin{equation}\label{eqt}
\sigma(q) \partial_q^2 t-\frac{1}{2} \sigma^{\prime\prime} (q) v^2 \partial_v^2 t+2\sigma^{\prime} (q) \partial_q t-2\sigma^{\prime\prime} (q) v \partial_v t=0.
\end{equation}
Now it is a standard matter to see that the system is hyperbolic, elliptic or parabolic if $\sigma^{\prime\prime}(q)$ is positive, negative
or zero, respectively \cite{Sommerfeld}.   For the KMP model one has $\sigma(q)=2 q^2$, and the inviscid MFT equations are hyperbolic.

\section{Calculation of $M$, $V$, $\cal H$ and $\dot{s}$ for the coupled-pulse solution}
\label{ab}
\subsection{Calculation of $M$ and $V$}
\label{ab1}

The main contribution to $M(t)$ comes from the $q$ pulse. Plugging Eq.~(\ref{qpulse1a}) in Eq.~(\ref{three}), we obtain
\begin{equation}\label{calcM}
    M(t)=\int_{-\infty}^{\infty} d\xi\, q_m\, \cosh^{-2} \frac{c(\xi-\xi_0)}{2} = \frac{4 q_m}{c}=\frac{1}{v_1}.
\end{equation}
Similarly, the main contribution to $V(t)$ comes from the $v$ pulse. Plugging Eq.~(\ref{vpulse1b}) in Eq.~(\ref{three}) yields
\begin{equation}\label{calcV}
    V(t)=\int_{-\infty}^{\infty} d\xi\, v_m\, \cosh^{-2} \frac{c\xi}{2} = \frac{4 v_m}{c}=\frac{1}{q_2},
\end{equation}

\subsection{Calculation of $\cal H$}
\label{ab2}

Plugging Eq.~(\ref{Q010}) in Eq.~(\ref{three}), we can write
\begin{equation}
  {\cal H}(t)=\int\limits_{-\infty}^{\infty} d\xi\,\left(-vq^{\prime}+q^2v^2\right) =\int\limits_{-\infty}^{\xi_1}  d\xi \,\left(-v_< q^{\prime}_{<}+q_<^2v_<^2\right)
  +\int\limits_{\xi_1}^{\infty}  d\xi\,\left(-v_> q^{\prime}_{>}+q_>^2v_>^2\right),
  \label{H010}
\end{equation}
where $q_<(\xi)$ and $v_<(\xi)$ are given by Eqs.~(\ref{vpulse1a})-(\ref{vpulse1c}),
$q_>(\xi)$ and $v_>(\xi)$ are given by Eqs.~(\ref{qpulse1a})-(\ref{qpulse1c}), and
$\xi_1>0$ obeys the strong inequality $\min \{c\xi_1,\,c(\xi_0-\xi_1)\}\gg 1$.
The first integrand on the right hand side of Eq.~(\ref{H010}) vanishes on the pulse solution, the second integrand is identically equal to $c v_1 q_>(\xi)$. Extending the lower integration limit in the second integral to $-\infty$, we obtain
\begin{equation}\label{calcH}
    {\cal H} (t)=\int_{-\infty}^{\infty} d\xi\, c v_1\, q_m \cosh^{-2} \frac{c(\xi-\xi_0)}{2} = c.
\end{equation}

\subsection{Calculation of $\dot{s}$}
\label{ab3}

Again, we write
\begin{equation}
 \dot{s}(t)=\int\limits_{-\infty}^{\infty} d\xi\,q^2v^2 =\int\limits_{-\infty}^{\xi_1}  d\xi \,q_<^2v_<^2+\int\limits_{\xi_1}^{\infty}  d\xi\,q_>^2v_>^2.
  \label{calcsdot}
\end{equation}
For the pulse solutions we have $q_<^2 v_<^2 = c v_< (q_<-q_2)$ and
$q_>^2 v_>^2 = c q_> (v_>-v_1)$. Then a straightforward integration, using Eqs.~(\ref{vpulse1a})-(\ref{qpulse1c}), gives in the leading order in $c\xi_1\gg 1$ and $c(\xi_0-\xi_1)\gg 1$
\begin{equation}
\int\limits_{-\infty}^{\xi_1}  d\xi \,q_<^2v_<^2 \simeq c^2 \xi_1
  \label{left10}
\end{equation}
and
\begin{equation}
\int\limits_{\xi_1}^{\infty}  d\xi \,q_>^2v_>^2 \simeq c^2 (\xi_0-\xi_1).
  \label{right10}
\end{equation}
Summing up the results of (\ref{left10}) and (\ref{right10}), we obtain $\dot{s}=c^2 \xi_0$, independent of $\xi_1$. This leads to Eq.~(\ref{PM234}).

\section{Finding $k_1(c)$}
\label{ac}

Let $k_1(c)=c \kappa(c)$. Then Eqs.~(\ref{J1}) and (\ref{A030}) become
\begin{equation}
   s(c,\Lambda)=\frac{e^{\Lambda}}{c \kappa(c)}-c \ln \kappa(c),\;\;\;j(c,\Lambda)=\frac{e^{\Lambda}}{\Lambda c \kappa(c)}.
\label{B1}
\end{equation}
The quantities $c$ and $\Lambda$ are related by Eq.~(\ref{delta2}) which we rewrite as $\phi(c,\Lambda)=0$, where we have defined
\begin{equation}
   \phi(c,\Lambda)=c^2 \kappa(c)-\text{Ei}(\Lambda).
\label{B2}
\end{equation}
Now we use Eqs.~(\ref{B1}) and (\ref{B2}) to calculate
\begin{equation*}
 \frac{d s}{d j}-\Lambda =\frac{\left|\frac{\partial(s,\phi)}{\partial(c,\Lambda)}\right|}
{\left|\frac{\partial(j,\phi)}{\partial(c,\Lambda)}\right|}-\Lambda
  =-\frac{c^2 \Lambda \, \kappa^2 (c) \left[\ln \kappa (c)-2\right]}{c^2
   (\Lambda -1) \,\kappa (c) \left[c \kappa^{\prime}(c)+2 \kappa
   (c)\right]-e^{\Lambda} \left[c \kappa^{\prime}(c)+\kappa (c)\right]}.
\end{equation*}
Setting this to zero yields $\ln \kappa(c)-2=0$, so $\kappa(c)=e^2$, independent of $c$.


\begin{thebibliography} {99}
\bibitem{4n2} A different convention, $\sigma(n)=4 n^2$, is often used for the KMP model. One can go
over from one convention to the other by rescaling $n$.
\end{thebibliography}
\end{document}